\documentclass[12pt]{article}
\usepackage{times}
\usepackage{geometry}
\usepackage[utf8]{inputenc}
\usepackage{enumitem,amssymb}
\usepackage[most]{tcolorbox}
\usepackage{ragged2e}
\usepackage{graphicx}
\usepackage{amsmath}
\usepackage{xcolor}
\usepackage[numbers,sort&compress]{natbib}
\usepackage{wrapfig}
\usepackage{floatrow}
\usepackage{booktabs}
\usepackage{multirow}
\usepackage{threeparttable}
\usepackage{parskip}
\usepackage{titlesec}
\titlespacing*{\section}{0pt}{2pt plus 1pt minus 1pt}{1pt}
\titlespacing*{\subsection}{0pt}{2pt plus 1pt minus 1pt}{0.5pt}
\titlespacing*{\subsubsection}{0pt}{1pt plus 1pt minus 1pt}{0.5em}
\titleformat{\subsection}[runin]{\normalfont\bfseries}{}{0em}{}[~\quad]
\titleformat{\subsubsection}[runin]{\normalfont\itshape}{\thesubsubsection}{0em}{}[\quad]

\usepackage{hyperref}

\DeclareRobustCommand{\ion}[2]{%
  \relax\ifmmode\text{#1\,\textsc{\lowercase{#2}}}%
  \else#1\,\textsc{\lowercase{#2}}\fi}

\newcommand{\hi}{\ion{H}{i}}
\newcommand{\heii}{\ion{He}{ii}}
\newcommand{\oi}{\ion{O}{i}}

\newcommand{\nii}{\ion{N}{ii}}

\newcommand{\nv}{\ion{N}{v}}
\newcommand{\oiv}{\ion{O}{iv}}
\newcommand{\ov}{\ion{O}{v}}
\newcommand{\ovi}{\ion{O}{vi}}
\newcommand{\neii}{\ion{Ne}{ii}}

\newcommand{\neviii}{\ion{Ne}{viii}}
\newcommand{\mgx}{\ion{Mg}{x}}
\newcommand{\svi}{\ion{S}{vi}}

\begin{document}
\thispagestyle{empty}
\RaggedRight

{\large Building a Roadmap for Hubble Science into the 2030s}\\[0.3em]
{\LARGE\textbf{Closing the UV Gap: Rest-frame EUV science from high-redshift QSOs as a legacy-defining capability}}\\[0.3em]
\normalsize\normalfont
\noindent\textbf{Authors:}
Rongmon Bordoloi$^{1}$ \& J. Michael Shull$^{2}$

\noindent{\small
$^{1}$NC State University; \texttt{\scriptsize rbordol@ncsu.edu},
$^{2}$University of Colorado Boulder; \texttt{\scriptsize michael.shull@colorado.edu}
}

\justifying
\begin{tcolorbox}[
    colback=blue!5!white,
    colframe=blue!50!black,
    title=\textbf{Executive Summary},
    fonttitle=\bfseries\large,
    fontupper=\normalfont\justifying,
    boxrule=0.8pt,
    left=2mm, right=2mm,
    top=1mm, bottom=1mm,
]
The Hubble Space Telescope is the only high-resolution ultraviolet spectroscopic facility that will exist until the Habitable Worlds Observatory (HWO) achieves first light in the mid-2040s.  We describe a coherent class of science, coupling rest-frame extreme-ultraviolet (EUV; 1--4~Ryd, 228--912~\AA) absorption and continuum spectroscopy of intermediate-redshift quasars at $z = 1-2$, shifting the rest-frame EUV photons into the HST/COS far-UV bandpass.  This science on quasars and gas in the IGM and CGM is doubly perishable.  The COS detector 
sensitivity is declining, just as new quasars are found (Milliquas, UVQS, and soon Rubin, Roman, and Euclid).  Thus, the window to reach  UV-bright quasars at $z>1$ QSOs narrows with every deferred orbit. Expanding HST UV orbit allocations in the 2030s would deliver a step-change in warm-hot CGM/IGM science and produce the first systematic, empirical EUV SED census of QSOs. These datasets will serve as the foundational low-redshift anchor for HWO science. This recommendation makes the scientific and strategic case for an expansion of the HST/COS spectroscopic data base on intermediate redshift AGN in their rest-frame EUV. %
\end{tcolorbox}

\setcounter{page}{1}

\section*{1.\ The Coming UV Gap}
\label{sec:gap}

A graduate student beginning a PhD in 2026 will likely be in their 
mid-40s before another high-resolution ultraviolet spectrograph is
operational in space.  Postdoctoral researchers and young astrophysics
faculty members will be in their 50s.  This is a planning fact, not a 
forecast.  It follows from the Astro2020 Decadal Survey's recommendation 
that HWO succeed HST as the flagship UV/optical capability \citep{astro2020},
with first light no earlier than the mid-2040s, and from the absence
of any approved bridging mission.

\begin{figure}[!t]
    \centering
    \includegraphics[width=\textwidth]{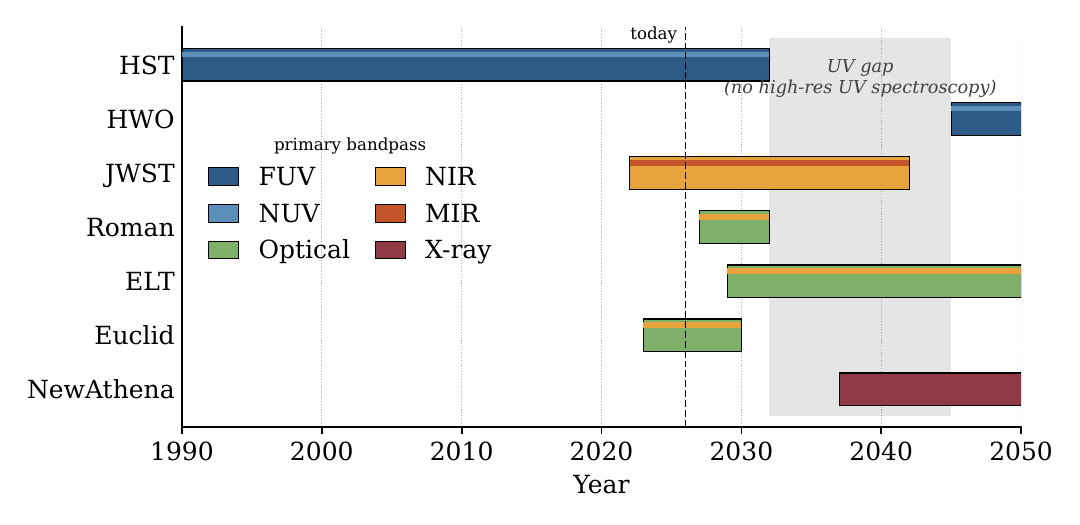}
    \caption{%
        UV/optical/infrared facility timeline from 1990--2050.  Colors denote
        primary spectroscopic bandpass (FUV, NUV, optical, NIR/MIR, X-ray).  
        The grey band marks the post-HST, pre-HWO interval during which no
        high-resolution UV spectroscopic capability exists.
    }
    \label{fig:timeline}
\end{figure}

The landscape leaves no ambiguity.  HST is the only operational telescope 
offering $R \gtrsim 10{,}000$--$20{,}000$ FUV spectroscopy below 
$\sim$3200~\AA.  JWST covers $0.6$--$28~\mu$m with no UV channel.  
Roman is an optical/NIR wide-field imager with no high-resolution spectrograph.  
CASTOR \citep{cote2019} and UVEX \citep{uvex} are UV imagers with low-resolution spectroscopy; these are  
not HST/COS successors.  New Athena cannot reach the rest-frame EUV at the resolving power needed for CGM/IGM diagnostics, and no ESA, JAXA, or CSA UV spectroscopic mission is in development.  Because Astro2020 did not recommend any bridging UV capability, high-resolution FUV spectroscopy will disappear once HST ceases science operations. It will not return for roughly two decades. Figure~\ref{fig:timeline} illustrates the gap.

\section*{2.\ Science enabled by the Rest-Frame EUV}
\label{sec:euv_science}

The extreme-ultraviolet (EUV) lies between 1--4~Ryd (wavelengths $\sim$228--912~\AA) spanning the spectral band between the \hi\ and \heii\ ionization edges.  Beyond the local interstellar medium (ISM), Galactic \hi\ absorption renders the local universe opaque in an ``EUV Dark Zone". However, rest-frame EUV observations are possible toward intermediate-redshift ($z \approx 0.5-2.1$) quasars and other active galactic nuclei (AGN), whose EUV photons are redshifted into the far-ultraviolet (FUV).  The required combination of a UV space telescope, high-redshift background sources, and spectral resolution sufficient to resolve narrow absorbers is satisfied by HST/COS alone.  This enables an entire new class of science.

Specific examples include the \textit{Warm-hot CGM/IGM diagnostics} using resonance absorption lines of \mgx~$\lambda\lambda$609, 624, \ov~$\lambda$630,  
\oiv~$\lambda$788, \neviii~$\lambda\lambda$770, 780, and 
\svi~$\lambda\lambda$933, 944 to detect gas at $T \sim 10^{5.5}$--$10^{6}$~K.  The \mgx\ and \neviii\ doublets trace $\sim10^{6}$~K plasma, gas otherwise accessible only through X-ray emission.  However, these absorption lines can
be detected at the high spectral resolution of HST/COS, resolving the line profiles and providing the kinematics and column densities that X-ray spectroscopy cannot. The intrinsic ionizing spectrum of AGN between 1--5~Ryd largely determines the photoionization corrections for CGM/IGM metallicities. Recent observations \citep{tilton2016,shull2012} demonstrate that 
per-object EUV slope measurements are now feasible, with some spectral indices ($\alpha_\nu \approx 1.0$--$1.1$) harder than the mean spectral index
$\alpha_{\nu} = 1.41\pm 0.15$ found with HST/COS QSO observations of 159 AGN \citep{stevans2014}.  The \heii\ Lyman-alpha forest traces \heii\ reionization at $z \sim 3$ \citep{shull2010, worseck2016, worseck2019}.
Models of the metagalactic ionizing background
\citep{haardt2012,khaire2019,puchwein2019,faucher2020} disagree by factors of several in the EUV and can be empirically anchored only with far-UV spectroscopy by HST.  No other facility reaches this regime: JWST is too red, ground-based telescopes are blocked by the atmosphere, 
and X-ray observatories cannot access the EUV longward of 100~\AA.  

\begin{figure}[!t]
    \centering
    \includegraphics[width=\columnwidth]{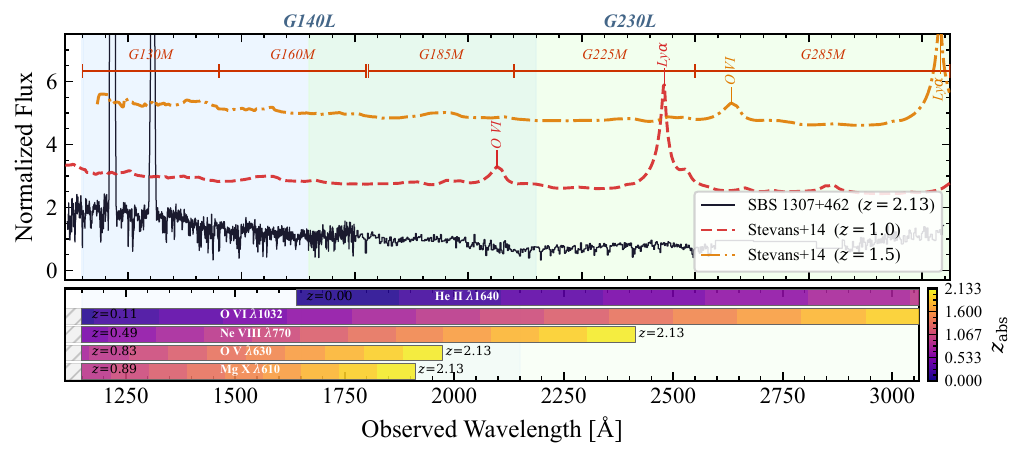}
    \caption{%
        Rest-frame EUV diagnostics in the COS bandpass.
        \textit{Top:} a UV-bright QSO ($z=2.13)$ observed with HST/COS 
        and two HST/COS QSO composite spectra shifted to $z=1.0$ and $z=1.5$, 
        respectively \citep{stevans2014}.  Several COS grating footprints 
        are overlaid.
        \textit{Bottom:} redshift ranges over which \mgx~$\lambda$610,
        \ov~$\lambda$630, \neviii~$\lambda$770, \ovi~$\lambda$1032,
        and \heii~$\lambda$1640 fall within the COS bandpass.
    }
    \label{fig:qso_sed}
\end{figure}

Figure~\ref{fig:qso_sed} illustrates the mapping in practice, with key EUV transitions moving into the HST/COS bandpass as a function of redshift. Over $z \sim 0$--$2$ a suite of diagnostic ions becomes accessible.

\section*{3.\ One EUV Program with Two Coupled Science Goals}
\label{sec:exhibits}

The science goals discussed below incorporate a single strategy: deep COS spectroscopy of UV-bright quasars at intermediate redshift, many with intervening metal absorption systems.  Each sight line probes warm-hot halo gas in EUV metal-line absorption \textit{and} characterizes the QSO ionizing spectral energy distribution (SED). The $z \sim 1$--$2$ interval is the sweet spot for AGN observations \citep{shull2012}: high enough to redshift the rest-frame EUV transitions into the COS bandpass, yet low enough that intergalactic Ly$\alpha$ forest absorption does not blanket the continuum. 

UV-bright QSOs at $z > 1$ are intrinsically scarce, owing to the presence of optically thick Lyman limit systems (LLS) along the sight lines. Earlier samples were also limited by the available all-sky UV photometry.  That situation has changed, with the Million Quasar Catalog matched to archival GALEX FUV photometry \citep[Milliquas;][]{milliquas}, the UV-Bright QSO Survey \citep[UVQS;][]{uvqs}, and Plane Quasar Survey \citep{planeqso} providing a large number of confirmed $z > 1$ targets. These AGN are reachable by deep COS spectroscopy at practical orbit cost, and the pool will 
grow as Rubin \citep{rubin2019}, Euclid \citep{euclid2022},and Roman 
\citep{roman2015} identify hundreds of new QSOs accessible to HST/COS.  

\subsection{Goal 1 --- Warm-Hot Halo Gas via EUV Metal Lines.}
\label{sec:nevi}

The warm-hot CGM/IGM ($T \sim 10^{5.5}$--$10^{6}$~K) holds the
dominant baryon reservoir in $L^{*}$ (or higher mass) halos and the warm-hot IGM 
\citep{savage2005,danforth2005,savage2011,mulchaey2009,Tumlinson2011,Bordoloi2014}.  Its standard tracer, the 
\ovi\ $\lambda\lambda1032,1038$ doublet, is often ambiguous. A given \ovi\ absorber could arise in  collisionally ionized gas near $10^{5.5}$~K or in cooler photoionized gas.  The EUV band can resolve this.  At $z \geq 1$, a single COS spectrum captures a large ensemble of ionic transitions of the same element \citep{verner1994} --- from \oi\ through \ovi, \nii\ through \nv, \neii\ through \neviii. That ion ladder, rather than any one line, determines both the temperature and ionization parameter.  The high-ionization doublets 
\neviii~$\lambda\lambda$770, 780 and \mgx~$\lambda\lambda$609, 624 extend the reach to gas at and above $10^{6}$~K, near the virial temperature of $L^{*}$ galaxy halos \citep{bordoloi2017}.

Exploiting these rich atomic transitions requires high-S/N spectroscopy with $R\sim 10,000 - 20,000$ (HST/COS G130M/G160M). The key EUV lines are weak, and only resolved profiles yield reliable equivalent widths and kinematics. Recent HST work \citep{tilton2016,shull2025} used low-resolution (COS G140L) spectroscopy on two $z>1.5$ UV-bright quasars with intervening absorbers (at $z\sim 1$) with $\log N_\mathrm{HI} \approx 17.2$--$17.4$ to recover the rest-frame EUV continuum down to $\sim$450~\AA. Such targets 
will require deep G130M/G160M follow-up to observe critical EUV lines (e.g., \mgx, \neviii) to complete the ionization census. To date, only $\sim$30--40 \neviii\ absorbers are confirmed in the literature \citep{tripp2011,meiring2013,hussain2015,bordoloi2017,burchett2019,CUBSNeVIII}. The sample is sightline-limited, not instrument-limited.

\subsection{Goal 2 --- QSO EUV Continuum Slopes.}
\label{sec:sed}

The strength and shape of the QSO ionizing continuum between 1--4~Ryd is the largest systematic uncertainty in CGM/IGM ionization modeling. The four most-used UV background models 
\citep{haardt2012,faucher2020,khaire2019,puchwein2019} disagree by factors of several in EUV emissivity.  Empirical constraints rest almost entirely on stacked composites 
\citep{telfer2002,shull2012,stevans2014,lusso2015}, which average over many objects and are biased toward bright unobscured AGN at $z \lesssim 1.5$. These composite spectra do not capture the wide diversity of individual AGN spectral slopes.  Recent HST work \citep{shull2025} demonstrates that 
that per-object slopes are now measurable, with EUV indices 
$\alpha_\nu = 0.98$--$1.11$ for some ultraluminous quasars significantly harder than the composite mean ($\alpha_{\nu} = 1.41\pm0.15$).  This is consistent with Comptonized accretion-disk components seen in MHD simulations \citep{jiang2025}.  The SED diversity is real, but composite spectra average it away.

A reliable slope needs more than FUV coverage. Joint FUV, NUV, and optical fitting is what reveals continuum curvature near rest-frame 1000~\AA\ and 
allows intervening Ly$\alpha$-forest and LLS absorption to be corrected, 
rather than being mistaken for intrinsic steepening.  This matters beyond QSO physics.  Accretion-disk modeling often fixates on the X-ray output, yet the EUV (13.6--100~eV) carries far more of the ionizing luminosity, governs CGM/IGM photoionization, and remains poorly reproduced by current disk models \cite{slone2012,laor2014,lusso2016}.

\textit{The combined program.}  A Treasury program targeting the 20--30 brightest $z > 1$ catalog QSOs delivers both goals
simultaneously.  Low-resolution G140L+G230L spectra (5--8 orbits per target) measure the EUV continuum slope and identify the sight lines with intervening absorbers suitable for follow-up.  $R\sim 15,000-20,000$ G130M/G160M at S/N~$\approx$~15--20 (20--30 orbits per target) will
detect the full EUV ion ladder (\mgx, \neviii, \ov, \oiv, and
intermediate Ne, N, O stages) at $\geq 3\sigma$ for 30--50\% solar abundance at 20\% ionization fraction \citep{bordoloi2017}.  
A total investment of $\sim$400--600 orbits over five cycles is a natural HST Treasury scope.  The result will be the first statistically controlled \neviii/\mgx\ census at $z \sim 1$ and the first per-object QSO EUV SED library. This survey will replace assumed template slopes with measured ones across the AGN luminosity function.  Both datasets anchor HWO CGM science; neither will exist unless built with HST.

\section*{4.\ Perishable Targets, Perishable Instrument}
\label{sec:perishable}

The urgency for UV spectroscopy is double-sided. It separates a case 
for \textit{preserving} HST UV time from the importance of its 
\textit{expansion} in the next five years.  On the instrument side, 
COS FUV sensitivity declines every cycle from detector aging.  
The same target requires more integration 
time each year, so every orbit deferred is an orbit that buys less 
science when it is executed.  On the target side, the AGN--EUV program 
depends on QSOs bright enough and at the right redshift to shift the 
diagnostic ions into the COS bandpass.  A selective AGN subset must 
be vetted with UV photometry before it can be prioritized. The brightest, 
most valuable sightlines are already identified, and the per-orbit yield of 
this program is highest today.  The UV spectroscopic capability has a 
closing window, not one the community can set aside and return to.

\section*{5.\ What Boosted Orbits Buy, and Legacy to HWO}
\label{sec:legacy}

A factor of 2--3 increase in UV absorption-line orbit allocations
sustained over five HST cycles produces a step change, not an
incremental one.  The \neviii\ absorber census would rise from $\sim$35
heterogeneous detections to a statistically controlled sample of
$\sim$150--300 systems.  More importantly, the deeper spectra push
beyond \neviii\ into the still-rarer high-ionization tracers ---
\mgx~$\lambda\lambda$609, 624 and other $\gtrsim$$10^{6}$~K diagnostics 
that are currently detected in only a handful of sight lines. The 
proposed survey will yield the first absorber sample large enough 
to map the hottest CGM/IGM phase rather than merely confirm its existence. 
The QSO EUV SED compilation moves from stacked spectral composites to 
resolved SEDs for $\gtrsim$100 targets spanning the AGN luminosity function.

These datasets are the foundational low-redshift anchor for HWO programs on
CGM, IGM, and reionization.  They will also be key observations for 
cross-checking with AGN spectra taken by UVEX \cite{uvex}.  
HWO is designed to push the absorption-line frontier to higher redshifts 
and lower column densities.  That science requires a well-characterized 
baseline of AGN at $z \lesssim 2$. Without the HST legacy datasets, 
HWO will arrive in the mid-2040s with no empirical zero-point for CGM 
ionization-state evolution or QSO SED systematics. 

The community needs to make this reframing explicit: HST UV orbits in the 
2030s are not in competition with HWO planning.  They are a prerequisite 
for HWO science return, and should be justified, budgeted, 
and communicated as infrastructure investment for the next flagship.

\section*{6.\ Recommendations}
\label{sec:recs}

\begin{itemize}[leftmargin=1.5em, itemsep=1pt]

    \item STScI should designate a minimum fraction of remaining HST
          orbits for UV absorption-line science each cycle, with
          EUV-class sight-line programs flagged as a priority category.

    \item The TAC should solicit and prioritize Large/Treasury programs
          targeting the UV-bright $z > 1$ QSO sample for
          \neviii/\mgx\ surveys and EUV continuum-slope characterization.

    \item The HWO Project Office should formally incorporate the
          $z \sim 0.5$--$2.1$ absorber catalogs and QSO EUV SED census
          into the HWO Science Traceability Matrix as enabling
          precursor datasets.

    \item COS should be maintained as the highest-priority HST instrument.

    \item COS calibration, lifetime-position monitoring, and
          sensitivity-tracking programs should be preserved and
          accelerated to maximize per-orbit science yield.

    \item NASA and STScI should communicate HST UV orbit allocations
          explicitly as HWO science infrastructure.

\end{itemize}

\newpage

\bibliography{refs}

\end{document}